\documentclass[letterpaper,https://www.overleaf.com/1592823166gntggcyjnjwg 10 pt, conference]{ieeeconf}

\IEEEoverridecommandlockouts                              

\usepackage{amsmath}
\usepackage{amssymb}
\usepackage{xcolor}
\usepackage{xspace}
\usepackage{bm}
\usepackage{subcaption}
\usepackage[font=small,labelfont=bf]{caption}
\usepackage[11pt]{moresize}

\newcommand{\parlength}[1]{}
\newcommand{\omitted}[1]{}

\usepackage[absolute]{textpos}
\title{\LARGE \bf
LAMP: Large-Scale Autonomous Mapping and Positioning for Exploration of Perceptually-Degraded Subterranean Environments
}
\vspace{-10pt}
\author{Kamak Ebadi$^{1}$, Yun Chang$^{2}$, Matteo Palieri$^{1}$, Alex Stephens$^{1}$, Alex Hatteland$^{1}$, Eric Heiden$^{3}$, Abhishek Thakur$^{4}$\\ Nobuhiro Funabiki$^{1}$, Benjamin Morrell$^{1}$, Sally Wood$^{5}$, Luca Carlone$^{2}$, Ali-akbar Agha-mohammadi$^{1}$
\thanks{The research was carried out at the Jet Propulsion Laboratory - California Institute of Technology, under a contract with the National Aeronautics and Space Administration. This work was partially funded by the Defense Advanced Research Projects Agency (DARPA). \textcopyright  2020 California Institute of Technology. }
\thanks{$^{1}$K.\,Ebadi, M.\,Palieri, A.\,Stephens, A.\,Hatteland, N. Funabiki, B.\,Morrell, and A.\,Aghamohammadi  are with NASA Jet Propulsion Laboratory - California Institute of Technology, Pasadena, CA 91109.}
\thanks{$^{2}$Y.\,Chang and L.\,Carlone are with Department of Aeronautics and Astronautics, Massachusetts Institute of Technology, Cambridge, MA 02139.}
\thanks{$^{3}$E.\,Heiden is with Department of Computer Science, University of Southern California, Los Angeles, CA 90007.}%
\thanks{$^{4}$A.\,Thakur is with Aptiv, Troy, MI 48098.
}%
\thanks{$^{5}$S.\,Wood is with Department of Electrical and Computer Engineering, Santa Clara University, Santa Clara, CA 95053. 
}%
}

\usepackage{mathtools}
\usepackage{color}
\usepackage{graphicx}
\newcommand{\ph}[1]{\textbf{#1:} }

\newcommand{\myParagraph}[1]{{\bf #1.}}
\newcommand{\iPCM}{{ICM}\xspace}
\newcommand{\MT}{\bm{T}}
\newcommand{\lc}{\text{lc}}
\newcommand{\odom}{\text{odom}}

\definecolor{green(ncs)}{rgb}{0.0, 0.62, 0.42}

\begin{document}

\begin{textblock}{20}(1,1)\centering
\small Please cite the paper as: K. Ebadi, Y. Change, M. Palieri, A. Stephens, A. H. Hatteland, E. Heiden, A. Thakur, N. Funabiki, B. Morrell, S. Wood, \\ L. Carlone, A. Agha-mohammadi. ``LAMP: Large-Scale Autonomous Mapping and Positioning for Exploration of Perceptually-Degraded Subterranean Environments." IEEE International Conference on Robotics and Automation, 2020.
\end{textblock}

\maketitle
\thispagestyle{empty}
\pagestyle{empty}

\begin{abstract}
Simultaneous Localization and Mapping (SLAM) in large-scale, unknown, and complex subterranean environments is a challenging problem. 
Sensors must operate in off-nominal conditions; uneven and slippery terrains make wheel odometry inaccurate, while long corridors without salient features make exteroceptive sensing ambiguous and prone to drift; finally, spurious loop closures that are frequent in environments with repetitive appearance, such as tunnels and mines, could result in a significant distortion of the entire map.
These challenges are in stark contrast with the need to build highly-accurate 3D maps to support a wide variety of applications, ranging from disaster response to the exploration of underground extraterrestrial worlds.
This paper reports on the implementation and testing of a lidar-based multi-robot SLAM system developed in the context of the DARPA Subterranean Challenge. 
We present a system architecture to enhance subterranean operation, including an accurate lidar-based front-end, 
and a flexible and robust back-end that automatically rejects outlying loop closures.
We present an extensive evaluation in large-scale, challenging subterranean environments, including the results obtained in the Tunnel Circuit of the DARPA Subterranean Challenge.
Finally, we discuss potential improvements, limitations of the state of the art, and future research directions.
\end{abstract} 

\section{Introduction\parlength{(1.25 pages)}}
Over the past two decades, there has been a growing demand for autonomous unmanned exploration of diverse subterranean environments, from tunnels and urban underground environments to complex cave networks. 
This surge of interest is motivated by terrestrial applications
such as search and rescue, disaster response, and infrastructure
monitoring, as well as the unparalleled scientific opportunities
offered by exploration of other worlds. Lava tubes,
subterranean conduits found on the Moon and Mars, are of particular interest: sheltered from cosmic
radiation and temperature fluctuations, they provide ideal conditions
for the development for microbial life, as well as potential sites for habitats in future human space missions~\cite{Haruyama}.

Localization and mapping are key capabilities for autonomous underground operation. In typical applications, there is no prior map of the environment, and GPS cannot be relied on to localize the robots. 
Indeed, the ultimate goal of many autonomous exploration applications \emph{is} to obtain an accurate 3D map of the environment.
 Unfortunately,  most SLAM systems have inadequate performance when deployed in perceptually-degraded subterranean environments:
 sensors must operate in off-nominal conditions (poor illumination or lack-thereof, dust, water puddles and non-Lambertian surfaces) which render visual-SLAM approaches unreliable~\cite{Tardioli1}; 
 uneven and slippery terrains make wheel odometry inaccurate, while long, featureless corridors make lidar-based mapping prone to drift; 
 finally, perceptual aliasing, the presence of many similar-looking corridors and intersections, induces spurious loop closures that can degrade the mapping results. 
\begin{figure}[t!]
\centering
	\includegraphics[width=1.0\columnwidth, trim= 0mm 0mm 0mm 0mm, clip]{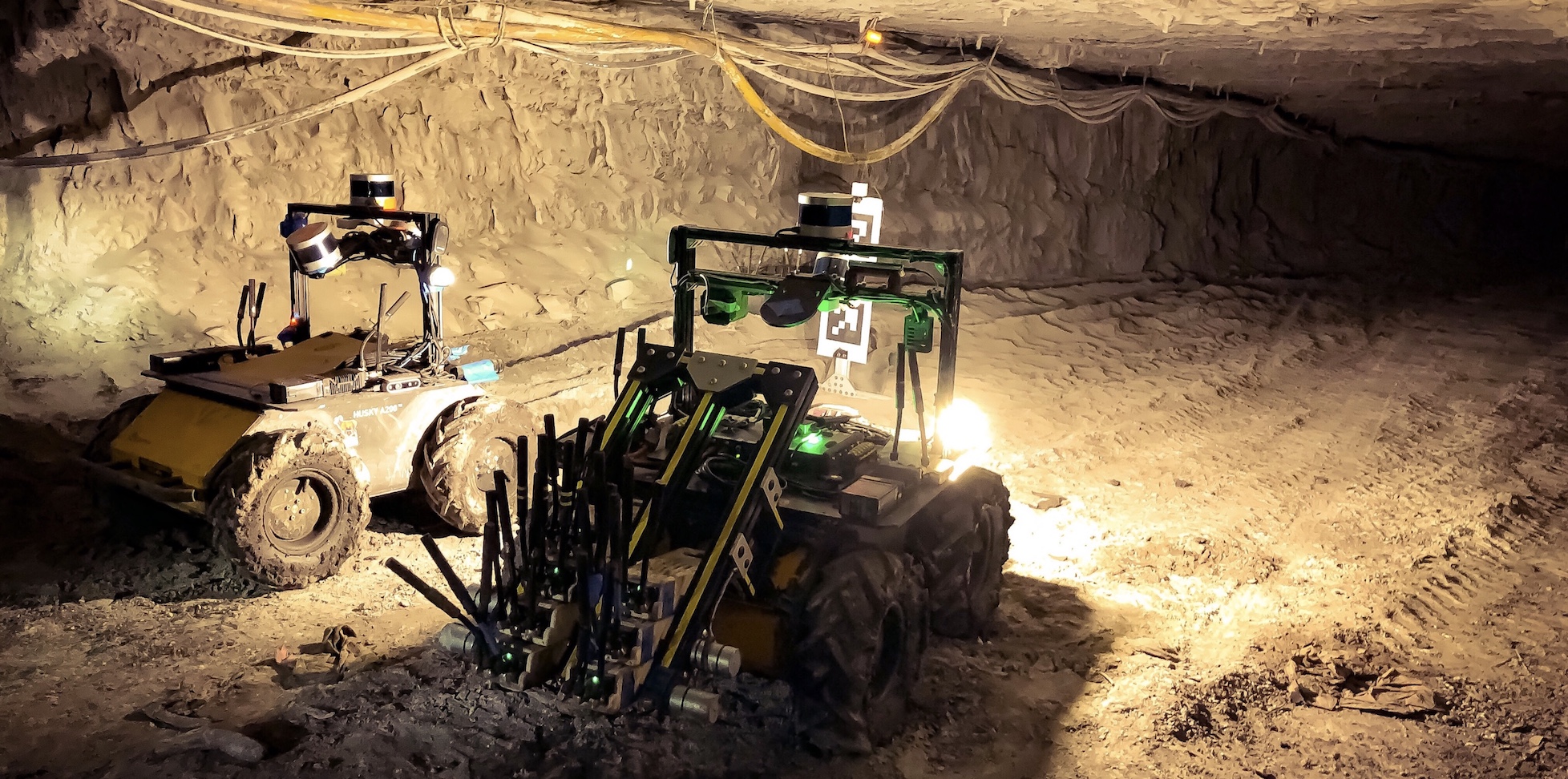} 
	\caption{Two autonomous ground rovers (Husky A200) explore the network of underground tunnels at the Arch mines. }
	\label{fig:cover} \vspace{-0.8cm}
\end{figure}

\myParagraph{Related work}
We review the literature on SLAM \emph{systems} for subterranean and perceptually degraded environments, and refer the reader to~\cite{Cadena} for a broader survey on SLAM.
Related work investigates different sensing modalities, including vision~\cite{ORB-SLAM}, visual-inertial~\cite{Bloesch, Leutenegger}, and 
thermal-inertial~\cite{Khattak1}. However, 
solely relying on vision for underground localization and mapping can be challenging as cameras are directional and sensitive to illumination changes and environmental conditions. Lidar sensors provide a $360^\circ$ horizontal field of view and high sampling rate, and do not rely on external light sources.
Therefore, 3D lidar-SLAM has been a popular solution to map complex unstructured environments, from early work~\cite{Nuchter} to more recent systems~\cite{LOAM, Cartographer, LeGO-LOAM, LLOAM, SegMap,Gentil}.
Thrun et al. \cite{Thrun} propose an underground mine mapping algorithm that relies on 2D scan matching and a global alignment step.
Tardioli et al.~\cite{Tardioli2, Tardioli3}  propose a system architecture for single and multi-robot exploration in underground tunnels.
Zlot et al.~\cite{Zlot} present a lidar-based SLAM method for mapping of a 17 km underground copper and gold mine;
to detect loop closures, they use a surfel representation and search for matches against previous trajectory segments.
Leingartner et al.~\cite{Leingartner} investigate the performance of off-the-shelf sensors and state-of-the-art mapping algorithms in mapping a 1.5 km long motorway tunnel in an urban search \& rescue scenario and conclude that the investigated sensing and mapping techniques are not yet robust enough to deal with these perceptually degraded environments.
Jacobson et al.~\cite{Jacobson} present a semi-supervised method that relies on manual selection of topological landmarks inside a 300 m long tunnel to perform localization using a low cost camera sensor. 

\myParagraph{Contribution}
This paper reports on the implementation and testing of a multi-robot SLAM system developed in the context of the Tunnel Circuit of the DARPA Subterranean Challenge (the first of the three subdomains)~\cite{SubT}, which aims to explore new approaches to rapidly map, navigate, search, and exploit large-scale complex underground environments.
The contribution of the paper is threefold.
First, we describe the multi-robot system architecture and the choice and implementation of the algorithmic modules. 
We comment on the improvements we made to existing algorithms to enhance subterranean operation, including 
  (i) an accurate lidar front-end based on scan-to-scan and scan-to-submap matching, 
  (ii) a versatile SLAM back-end that can incorporate data from multiple sensors and inputs from a human operator, 
   and  (iii) an Incremental Consistent Measurement (ICM) Set Maximization approach based on~\cite{PCM} to reject outlying loop closures.  
The second key contribution is an extensive evaluation in challenging underground mines across the United States, including experiments conducted in the Bruceton Research mine in August 2019, as part of the Tunnel Circuit of the DARPA Subterranean Challenge, where the system described in this paper has afforded the CoSTAR team~\cite{CoSTAR} the second place. 
Finally, we discuss the lessons learned, including potential improvements, limitations of the state of the art, and future research directions.
\section{LAMP: Large-scale Autonomous \\ Mapping and  Positioning \parlength{(2.25 pages)}}
\newcommand{\calX}{\mathcal{X}}
\newcommand{\calP}{\mathcal{P}}
\newcommand{\calG}{\mathcal{G}}

This section introduces our \emph{Large-scale Autonomous  Mapping  and  Positioning} (LAMP) system. 
Using data from 3D lidar scanners of one or more robots, LAMP estimates the trajectories of each robot as well as a point cloud map of the full environment. Moreover, it uses an RGB-D camera to detect and localize known objects in the environment; we call these objects \emph{artifacts} (following~\cite{SubT}) to distinguish them from 
 natural landmarks commonly used in landmark-based SLAM.
 While the system is designed to work with multiple robots (all connected to a central base station), it naturally falls back to a single-robot SLAM system when 
 the base station is out of communication range. 

\subsection{Overview: Single and Multi-robot Architecture \parlength{(0.5 pages)}}

\begin{figure}[t!]    
\centering 
  \includegraphics[width=1.0\columnwidth]{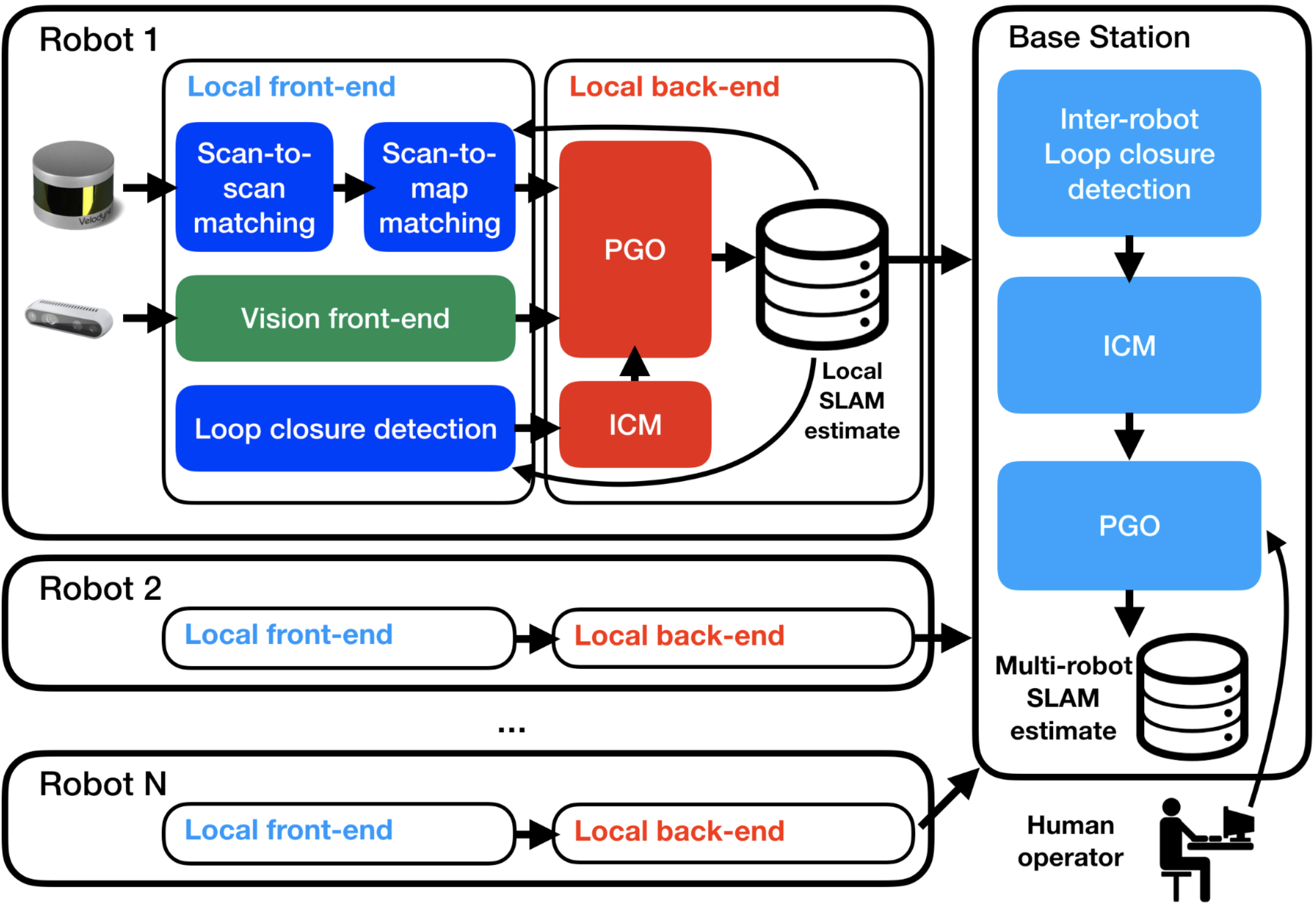}\vspace{-0.5cm}
  \caption{Centralized multi-robot SLAM architecture. \label{fig:multiArchitecture}\vspace{-0.6cm}}
\end{figure}

Fig.~\ref{fig:multiArchitecture} shows LAMP's multi-robot architecture.
The local front-end includes a lidar front-end (Section II-B, module shown in blue in Fig. 2) that computes odometry estimates and detects loop closures, and a vision front-end (Section  II-C,  module  shown  in  green  in Fig. 2) for artifact detection and localization. 
The local back-end uses pose graph optimization (PGO)~\cite{Cadena} to estimate the trajectory of the robot (Section~\ref{sec:PGO}) along with the location of artifacts, and also includes an outlier rejection mechanism, based on ICM (Section~\ref{sec:PCM}). 
For robot $i$, the result of the local front-end and back-end is an estimate of the robot trajectory (modeled as a sparse set of poses) and artifact positions, as well as a downsampled point cloud attached to each robot pose; we denote 
with $\calG_i$ and $\calX_i$ the pose graph and the pose estimates for robot $i$,
and we denote with $\calP_i$ the collections of local point clouds for robot $i$. 

When within communication range, each robot sends the tuple $(\calX_i,\calG_i,\calP_i)$ to a base station.
The base station fuses the estimates from all robots by searching for potential loop closures between the maps (inter-robot loop closure detection module), discarding outliers (\iPCM module), and solving a second PGO problem, this time involving all robot trajectories and 
all observed artifacts. We discuss the key modules from Fig.~\ref{fig:multiArchitecture} in the following sections.
\subsection{Local Lidar Front-end \parlength{(0.5 pages)}}
\label{sec:lidarFrontend}
The lidar front-end is in charge of using the raw lidar data to obtain relative pose measurements between consecutive robot poses (odometry) 
and non-consecutive poses (loop closures). 
The front-end builds upon an existing open-source implementation~\cite{BLAM} and proceeds along the same lines as modern lidar-SLAM systems such as 
LOAM~\cite{LOAM}.

\subsubsection{Lidar Odometry}\label{sec:lidarOdometry}
A typical 3D lidar, such as the Velodyne Puck LITE~\cite{PUCK}, generates up to 10,000 points per scan and produces 10 scans per second.
In order to increase efficiency and filter out noise, we apply a 3D voxel grid filter to the raw point cloud, using the implementation 
available in the \emph{Point Cloud Library} (PCL)~\cite{PCL}.
We then use the Generalized Iterative Closest Point (GICP) algorithm \cite{GICP} to perform \emph{scan-to-scan matching}, which estimates the relative 
motion between the point cloud collected at the current time $t$ and the point cloud collected at time $t-1$. 
We initialize the GICP algorithm with the identity pose.
Since compact lidars, such as the Velodyne Puck, have relatively narrow vertical field of view ($30^{\circ}$) and 
low vertical angular resolution ($2.0^{\circ}$, 16 scanning channels), the scan-to-scan estimates are not very accurate.
Therefore, as done in other lidar-SLAM pipelines~\cite{LOAM},
we also use a \emph{scan-to-submap matching} step, where the latest scan is matched against a local submap. 
Note that both matching modules produce an odometry estimate, but scan-to-submap matching typically produces more accurate estimates, as we show in the experimental section. The odometric estimate from the scan-to-submap matching is then passed to the back-end.
\subsubsection{Loop Closure Detection}\label{sec:lidarLC}

In order to find loop closures, we compare the current scan at time $t$ with scans in $\calP_i$ within a given radius (10 m in our tests), see Fig.~\ref{fig:ICM}.
The current scan is then registered against each of the nearby scans using GICP to obtain a transformation.
In order to find loop closures, GICP is initialized with zero translation and a rotation from the current pose estimates (from odometry). Zero translation was found to give more reliable results in narrow tunnel environments and avoid local minima in GICP, leveraging the topology of the environment that limits the differences in  possible positions.
After executing GICP, we check the fitness score~\cite{PCL} 
and reject loop closures with score larger than a threshold (5.0 in our experiments). 
The loop closures that pass the fitness test are then passed to the local back-end.
Since loop closure detection involves executing GICP  for many pairs of scans, we execute it on a parallel thread, and we only execute it at a 
subset of scans (the \emph{key-scans}, see Section~\ref{sec:PGO}). 

\subsection{Vision Front-end and Artifact Positioning}
\label{sec:visionFrontend}

The goal of the vision front-end is to detect objects of interest and estimate their relative position with respect to the robot. 
For instance, in the context of the DARPA Subterranean Challenge, the objects of interest are either \emph{artifacts} (known objects of interest), 
or \emph{fiducial markers} at known locations (only present at the entrance of the tunnel, for calibration purposes).
We use YOLO~\cite{YOLO} to detect the artifacts and the AprilTags library~\cite{Apriltag1, Apriltag2, Apriltag3} to detect the fiducial markers.
YOLO produces a 2D bounding box around a detected artifact; we then estimate the position of the artifact by using the bearing corresponding to the center of the bounding box and the range obtained from the depth channel of the RGB-D camera.
\subsection{Single and Multi-robot Pose Graph Back-end}
\label{sec:PGO}

We use a versatile SLAM back-end based on pose graph optimization (PGO)~\cite{Cadena} and implemented in GTSAM~\cite{GTSAM}. 
The same PGO implementation is executed both locally at each robot and at the base station, as discussed below.

\subsubsection{Local (Single-Robot) PGO}
The local back-end at robot $i$ receives the odometry measurements from the front-end and periodically instantiates a new key-scan: a key-scan is created after a minimum odometric displacement  ($30^\circ$ rotation or $1m$ translation) and the corresponding pose and odometry edge are added to the local pose graph $\calG_i$. 
When the back-end receives loop closure measurements from the Loop Closure Detection Module, we first check their quality using the 
\iPCM module 
described in Section~\ref{sec:PCM}, and reject outliers. Then, a loop closure edge is added to  $\calG_i$.
Finally, when the vision front-end produces an artifact (or fiducial) detection, a new variable is instantiated in the graph (the unknown artifact position) 
and an edge is added between the artifact and the corresponding observation pose. In the case of a second observation of an artifact, an additional edge is added between the observation pose and the existing artifact node. 
\omitted{For the Apriltag fiducial, we also add a prior on the fiducial markers pose, as its position in the world coordinate frame is known.}
To compute the trajectory estimate, we optimize the pose graph using the Gauss-Newton implementation in GTSAM.
\subsubsection{Base station (Multi-robot) PGO}
The base station receives the single pose graphs $\calG_i$ from each robot $i$ within communication range and merges them together in a single pose graph $\calG$. We assume that the robots share a common reference frame (in practice, the fiducial markers can be used to establish a common frame). Before optimizing the multi-robot graph $\calG$, we search for inter-robot loop closures;
since all robots share a common reference frame, this can be done using the same approach described in Section~\ref{sec:lidarLC}.

\emph{Human-in-the-loop Pose Graph Editing.} The versatility of our PGO also enables a more fluid interaction with the human operator. 
In some cases the operator can identify  (by inspecting the map)  a loop closure that has not been detected by the lidar front-end.
In our system, the operator can use rviz~\cite{RVIZ} and the Robot Operating System (ROS)~\cite{ROS} to interact with the map and attempt a loop closure between a pair of poses: this is done by calling a ROS service with the IDs of the potential loop closure poses, which results in the addition of an edge in the pose graph. This functionality can be used during operation or as a post-processing step.

\subsection{Incremental Consistent Measurement Set Maximization}
\label{sec:PCM}
Before executing PGO, we check the quality of the lidar loop closures and remove outliers.
This is done by the Incremental Consistency Maximization modules in Fig.~\ref{fig:multiArchitecture}. 
The ICM modules are largely based on the Pairwise Consistent Measurement (PCM) Set Maximization algorithm of Mangelson et al.~\cite{PCM}, 
which we modify to enable online (single and multi-robot) operation.
\iPCM includes the following steps.

\begin{figure}[h!]    
\centering
  \includegraphics[width=1\columnwidth, trim= 0mm 0mm 0mm 0mm, clip]{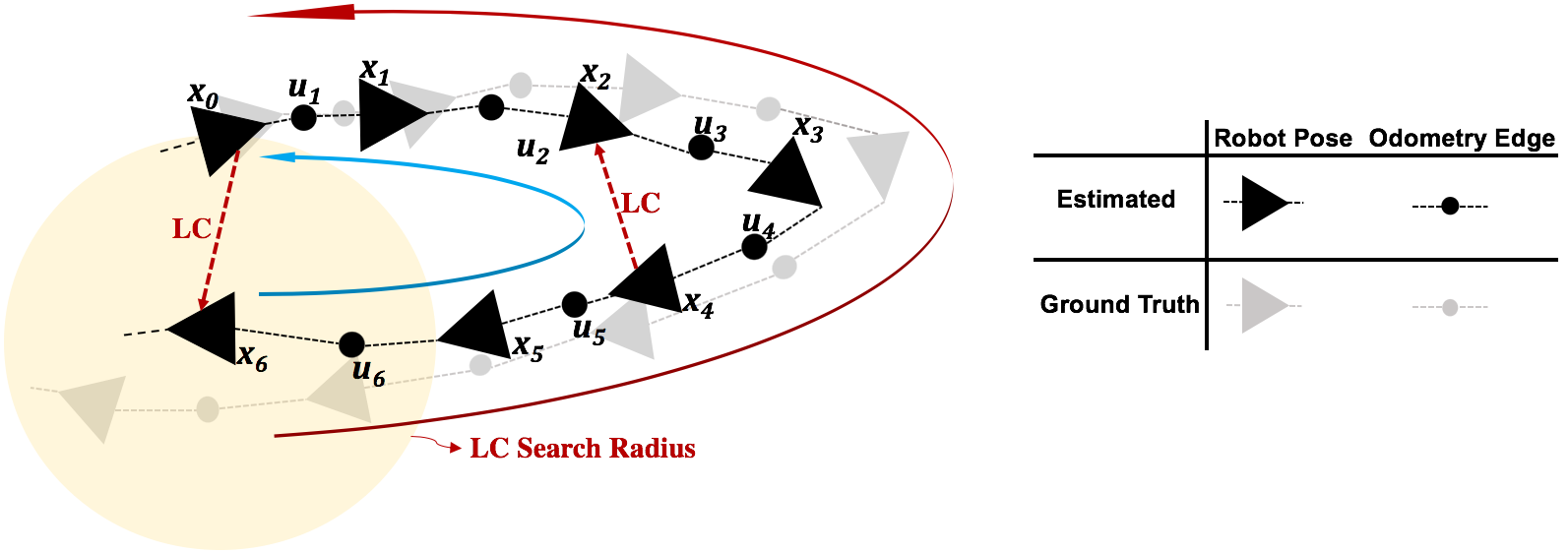}
  \caption{Local pose graph $\calG_i$: poses, edges, and cycles. \vspace{-3mm}} 
  \label{fig:ICM}
\end{figure}

\emph{Odometry Check.} While \iPCM has been designed for the multi-robot case, we adapt it to the single-robot case.
In the single-robot case, each loop closure can be tested both in terms of pairwise consistency against other loop closures  as in~\cite{PCM}, and against the odometry 
edges.
As in PCM, the basic observation is that, in the absence of noise, relative pose measurements should compose to the identity along cycles in the graph~\cite{Carlone14}. Cycles are defined as complete loops in the graph, as depicted in Fig.~\ref{fig:ICM}.
Assume that the front-end produces a putative loop closure measurement $\MT^{\lc}_{ij}$ of the relative pose between poses $\MT_i$ and $\MT_j$. 
Then, the measurements along the cycle formed by the loop closure  $\MT^{\lc}_{ij}$ and the odometry edges (in red in Fig.~\ref{fig:ICM}) must compose to the identity:
\begin{equation}\label{eq:aveErr1}
\MT^{err}_{ij} \doteq \MT^{\lc}_{ij} \cdot \MT^{\odom}_{ji}
\end{equation}
where $\MT^{\odom}_{ji}$ is the relative pose obtained by chaining the odometry edges from pose $j$ to pose $i$. 
Eq.~\eqref{eq:aveErr1} represents the pose error accumulated along the loop. 
If $\MT^{\lc}_{ij}$ is an outlier, it will cause the error $\MT^{err}_{ij}$
to be large. Denoting with $m$ the number of edges along the loop (7 in the example of Fig.~\ref{fig:ICM}), 
we reject as outliers loop closures inducing average errors $\frac{1}{m}\MT^{err}_{ij}$ larger than a threshold 
(in our tests, we limit the rotation error to 0.05~rad and the translation error to 0.1~m).

\emph{Pairwise Consistency Check.}
 Using a check similar to the one described in the previous section, we check pairwise consistency between pairs of loop closures.
Assume that the front-end produces a putative loop closure measurement $\MT^{\lc}_{ij}$. 
 In order to check if $\MT^{\lc}_{ij}$ is consistent with a previous loop closure $\MT^{\lc}_{kl}$, 
 we compute the loop error as:
 \begin{equation}\label{eq:aveErr2}
\MT^{err}_{ij,kl} \doteq \MT^{\lc}_{ij}\cdot \MT^{\odom}_{jl} \cdot \MT^{\lc}_{lk} \cdot \MT^{\odom}_{ki}
\end{equation}
and we build an adjacency matrix keeping track of which pair of loop closures are pairwise consistent, as in~\cite{PCM}.
Note that we use the simplified check~\eqref{eq:aveErr2} instead of the probabilistic check proposed in~\cite{PCM} since we do not have access to precise covariances for the odometry and loop closures. 

\emph{Incremental Consistency Maximization.}
So far we have rejected gross outliers using the odometry check and have built an adjacency matrix that keeps track of which pair of 
loop closures is consistent. 
In our implementation, the matrix is not built from scratch, but we only add a row and column when a new loop closure is detected and passes the odometry check. Then, we follow~\cite{PCM} and find a set of mutually consistent loop closures by computing the maximum clique on the adjacency matrix. Loop closures in the maximum clique are considered inliers and added to   the graph. 

\begin{table}[t]
\centering
\caption{Underground tunnels and mines explored in our field experiments.
The distance explored, width and height are averaged and rounded values from our experiments.}
\label{tab:mines}
\resizebox{\columnwidth}{!}{%
\begin{tabular}{|l|l|l|l|}
\hline
\textbf{Name}           & \begin{tabular}[c]{@{}l@{}}\textbf{Distance}\\\textbf{Explored}\end{tabular} & \textbf{Type} & \textbf{Location} \\ \hline
Arch Pocahontas Mine    & 1100 m             & Coal Mine             & Beckley, WV     \\ \hline
Beckley Exhibition Mine & 1000 m            & Coal Mine           & Beckley, WV              \\ \hline
Bruceton Safety Research (SR) Mine    & 1400 m             & Coal Mine       & Pittsburgh,  PA             \\ \hline
Bruceton Experimental (EX) Mine       & 700 m           & Coal Mine      & Pittsburgh,  PA             \\ \hline
Highland Mine           & 1400 m            & Coal Mine             & Logan, WV             \\ \hline
Eagle Mine       & 500 m             & Gold Mine       & Julian, CA              \\ \hline
\end{tabular}%
 }
 \vspace{-5mm}
\end{table}

\begin{figure}[t!]    
\centering 
  \includegraphics[width=1.0\columnwidth]{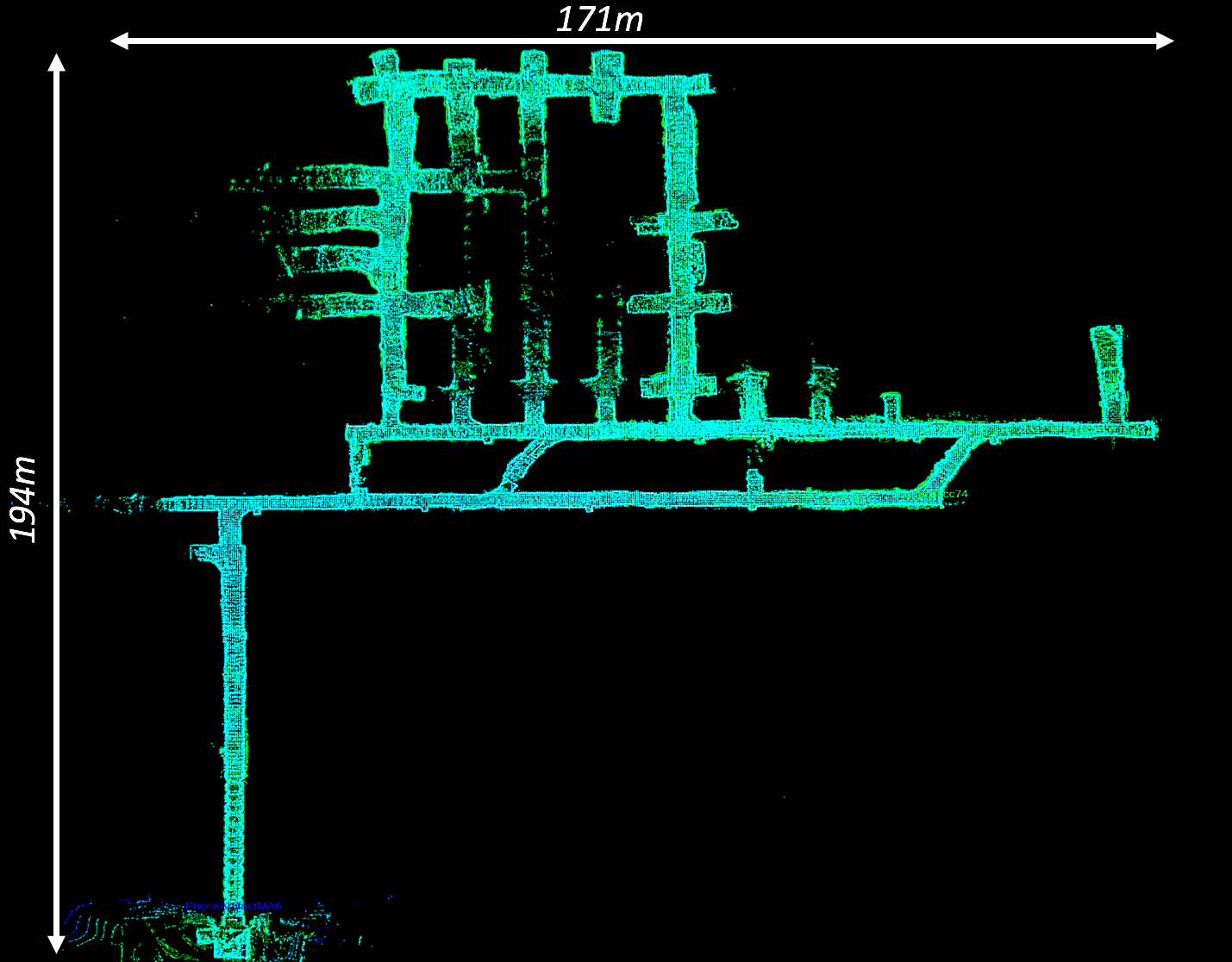}
  \caption{Partial ground truth map of the Bruceton Safety Research Mine, obtained by optimizing the best robot trajectory with ground truth position of fiducials, artifacts, and total station measurements during the tunnel circuit of the DARPA Subterranean Challenge. \vspace{-5mm}
  }
  \label{fig:SR-GT}
\end{figure}

\section{Experiments}

\subsection{Test Environments and Setup}
We perform experiments in six underground tunnels and mines, listed in Table \ref{tab:mines}. 
These environments are extremely challenging both for perception and mobility. 
In the following we provide a short description of our system setup.

\ph{Robots} The robots used in our tests are Husky-A200 series ground vehicles, equipped with VLP-16 Puck Lite lidar, an Intel RealSense D435 RGB-D camera, and an Intel NUC 7i7DNBE  ($4\times1.9$ GHz, 32 GB RAM) processor.
The  robots rely on communication radios (visible on the back of the robot in Fig.~\ref{fig:cover}) which create a mesh network to provide communication between robots and the base station.

\ph{Base Station} The base station is an Intel Hades Canyon NUC8i7HVKVA ($4\times1.9$ GHz, 32 GB RAM). The base station is placed outside the mine to receive incoming telemetry data as well as pose graphs, and artifact localization results from all the robots deployed in the tunnel.

\ph{Ground Truth} Obtaining ground truth trajectory and map estimates in large-scale underground environments is a challenging task. 
After the tests in the Bruceton mines, DARPA provided ground truth (surveyed) locations of the artifacts. To obtain a proxy for the ground truth trajectories of our robots, we enforced the ground truth locations of the artifacts in the pose graph of each robot, and used the resulting optimized trajectory as ground truth as presented in Fig. \ref{fig:SR-GT}; when available, we also added sporadic measurements of the robot poses obtained using a total station (TS)~\cite{total-station}. Since ground truth artifact locations are only available for the Bruceton mines, 
we report quantitative metrics for these mines, and visualize mapping results for the others.

\subsection{Results and Evaluation}

\begin{figure}[t!]   
\centering 
	\includegraphics[width=1.0\columnwidth]{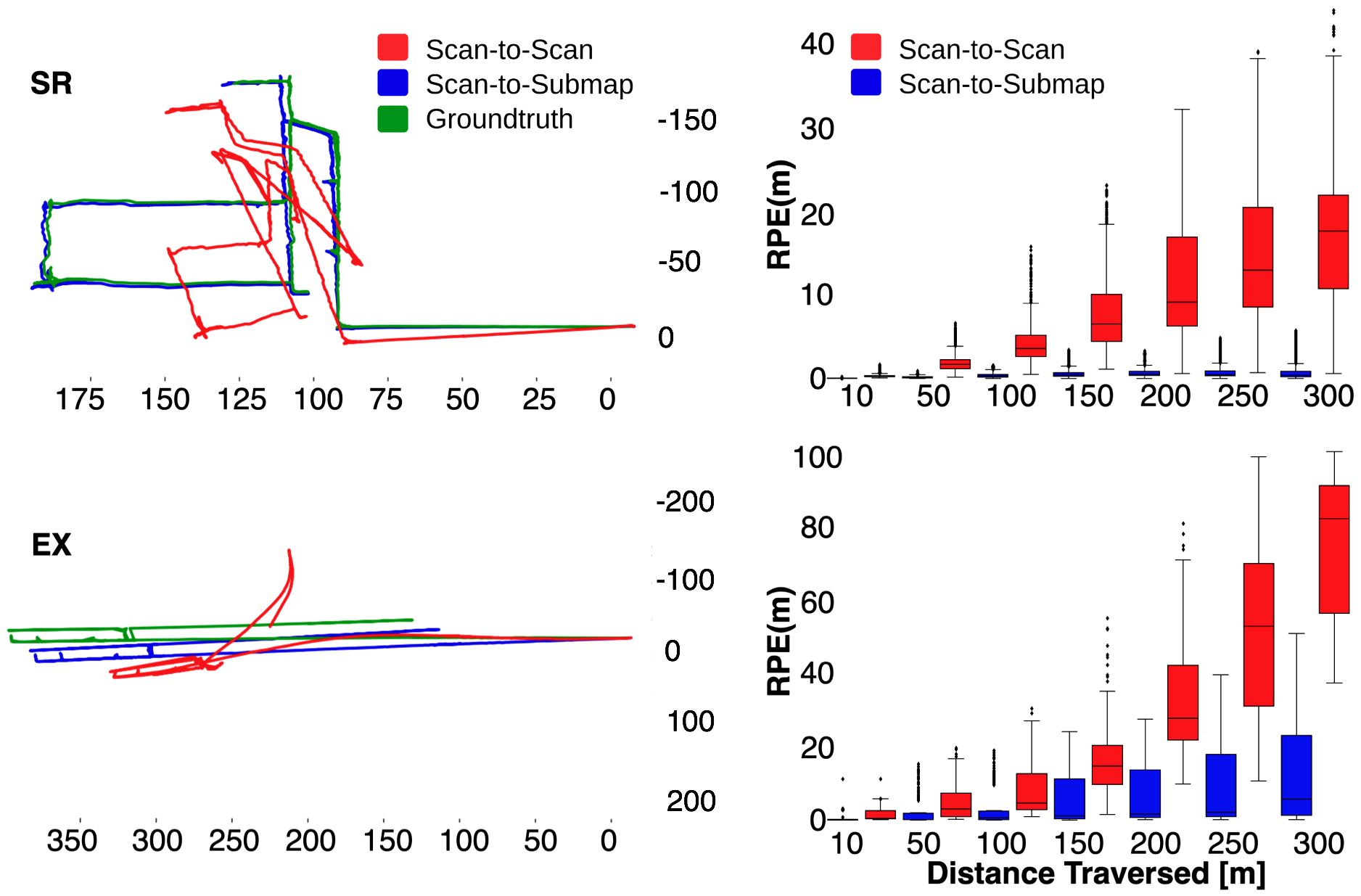}
	\caption{Left: odometric trajectories from scan-to-scan and scan-to-submap matching versus ground truth for 
  (a) Bruceton Safety Research (SR) mine and (b) Experimental (EX) mine. 
  Right: drift statistics for (c) SR and (d) EX.\vspace{-3mm}}
	\label{fig:frontend-drift-qual-eval}
\end{figure}

\subsubsection{Odometry Evaluation}
In this section, we show that the scan-to-submap matching (i) drastically improves the odometry accuracy when compared to the scan-to-scan matching, 
and (ii) in nominal conditions, it has a drift of $1\%$ of the distance traveled. 
Fig.~\ref{fig:frontend-drift-qual-eval} shows results in the Bruceton mines. 
Fig.~\ref{fig:frontend-drift-qual-eval}(a)-(b) report the trajectory estimates for the Safety Research and the Experimental mine, respectively.
The scan-to-scan matching quickly drifts, as it poorly estimates the pitch of the robot, due to the low vertical resolution of the Puck. The scan-to-submap matching has significantly better performance. 
Fig.~\ref{fig:frontend-drift-qual-eval}(c)-(d) provides a quantitative evaluation of robot drift in terms of Relative Pose Error (RPE) using \emph{evo}~\cite{Grupp}. 
In the Safety Research mine, the scan-to-submap odometry has a drift smaller than $1\%$ of the distance travel. On the Experimental mine, the drift is larger due to the presence of featureless corridors that cause GICP to underestimate the motion of the robot along the direction of the tunnel (see discussion in Section~\ref{sec:discussion}). 

\begin{figure}[t!]   
\centering 
  \includegraphics[width=1.0\columnwidth]{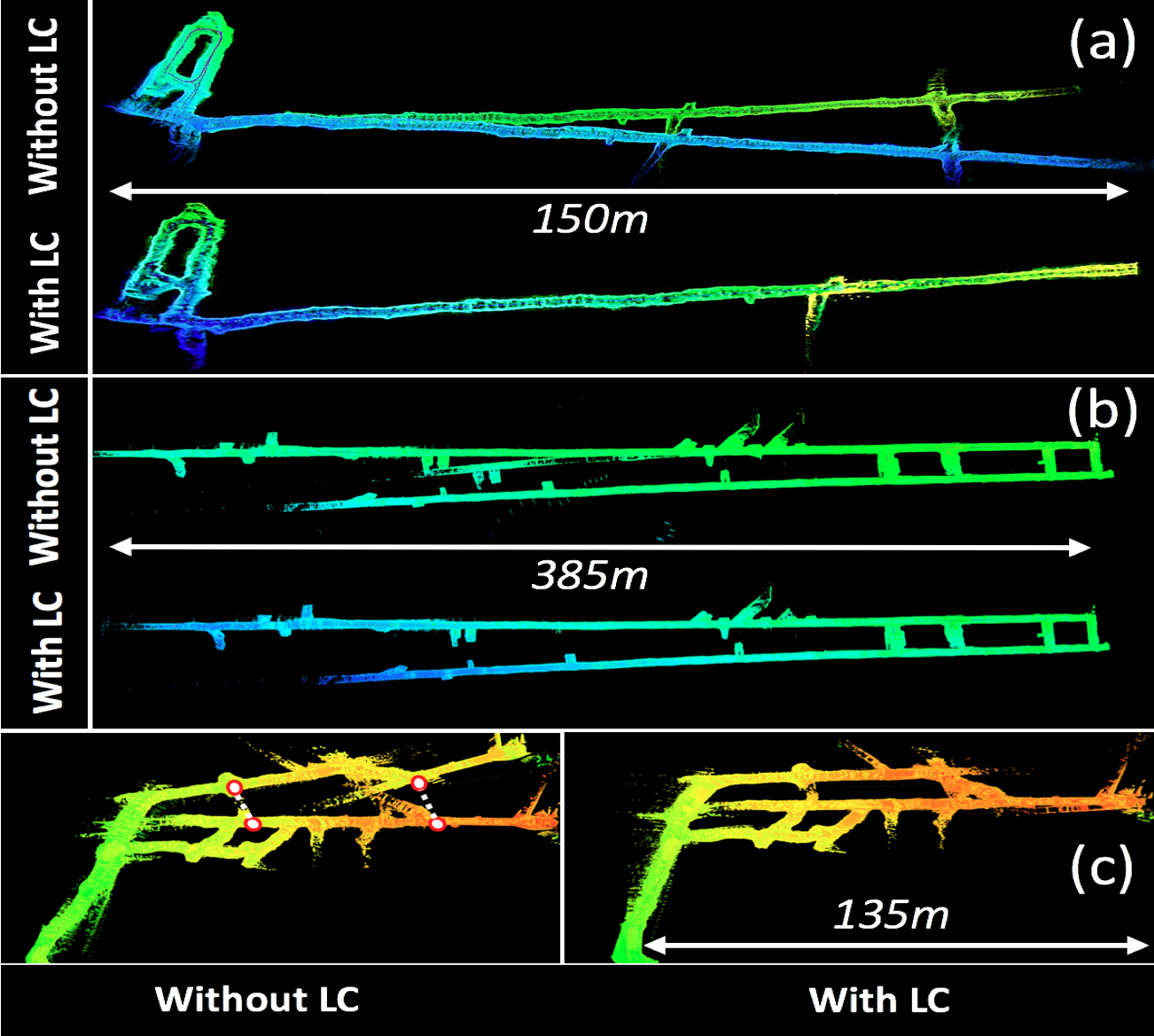}
  \caption{Map estimates with and without loop closure (LC) detection for (a) Eagle mine, (b) Bruceton Experimental Research mine, (c) Beckley Exhibition Coal Mine.\vspace{-4mm}\label{fig:loopclosure}}
\end{figure}
\begin{figure}[h!]    
\centering 
  \includegraphics[width=1.0\columnwidth]{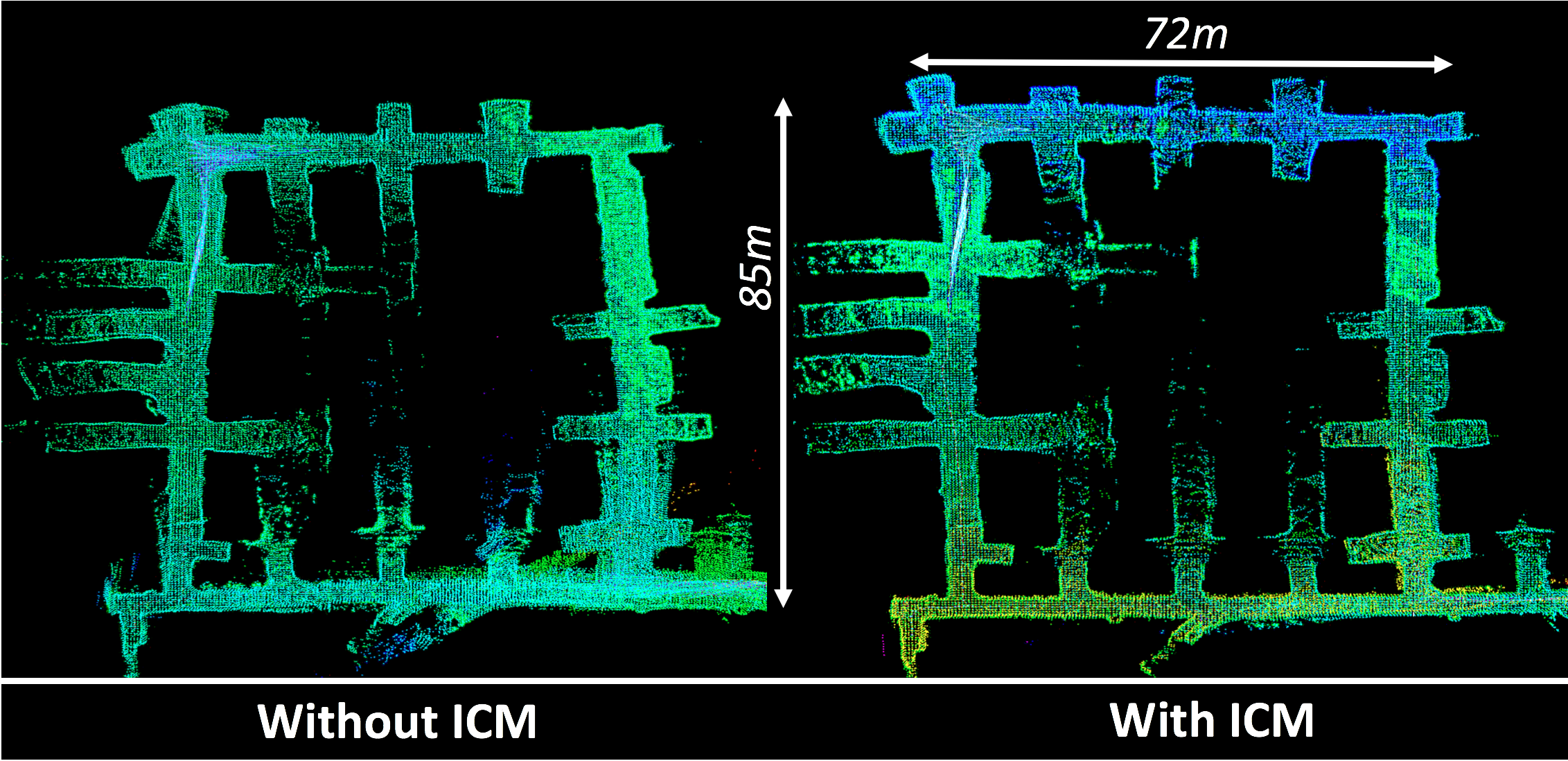}
  \caption{A section of Bruceton Safety Research mine. Left: PGO without \iPCM; the map is visibly distorted. 
  Right: PGO with \iPCM; map has no distortion since outlier loop closures are rejected. \label{fig:PCM}\vspace{-6mm}}
\end{figure}

\subsubsection{Loop Closure Evaluation}
Fig.~\ref{fig:loopclosure} reports a selected set of maps obtained with and without loop closure detection (Section~\ref{sec:lidarFrontend}) in three underground mines. 
Fig.~\ref{fig:loopclosure}(a)-(b) show that loop closure detection  avoids the appearance of spurious corridors in the map (resulting from the odometry drift) in (a) Eagle Mine and (b) Bruceton Experimental Mine. 
Fig.\ref{fig:loopclosure}(c) shows the map obtained in the Beckley Exhibition Coal mine. 
This is particularly interesting, due to the presence of a very long loop. In this case, using the standard (odometric) initialization in GCIP does not allow closing the loop; the loop closure poses highlighted in red in Fig.\ref{fig:loopclosure}(c) are too far, and GICP does not converge to a reasonable estimate. However, the approach in Section~\ref{sec:lidarFrontend} is still able to correctly detect and enforce the loop closure.

\subsubsection{\iPCM Evaluation} 
Spurious loop closures are frequent in environments with repetitive appearance, such as tunnels and mines. 
Including outliers in the PGO typically results in a significant distortion of the entire map, hence their rejection is crucial to obtain accurate maps.
Fig.~\ref{fig:PCM} shows the impact of \iPCM on the mapping results (more results will also presented in the multi-robot subsection).
The map on the left is obtained without \iPCM and is distorted due to inaccurate loop closures.
The map on the right is obtained by enabling \iPCM: poor loop closures are removed by \iPCM and the resulting final map shows no visible distortion. We evaluate the quality of loop closures by measuring the end-to-end drift where the robot is started and stopped at the same location in the tunnel. The end-to-end error in the map obtained with \iPCM is 0.2~m which is significantly less than the 2.8~m end-to-end drift in the map without \iPCM.

\begin{figure}[t!]
\centering 
	\includegraphics[width=1.0\columnwidth]{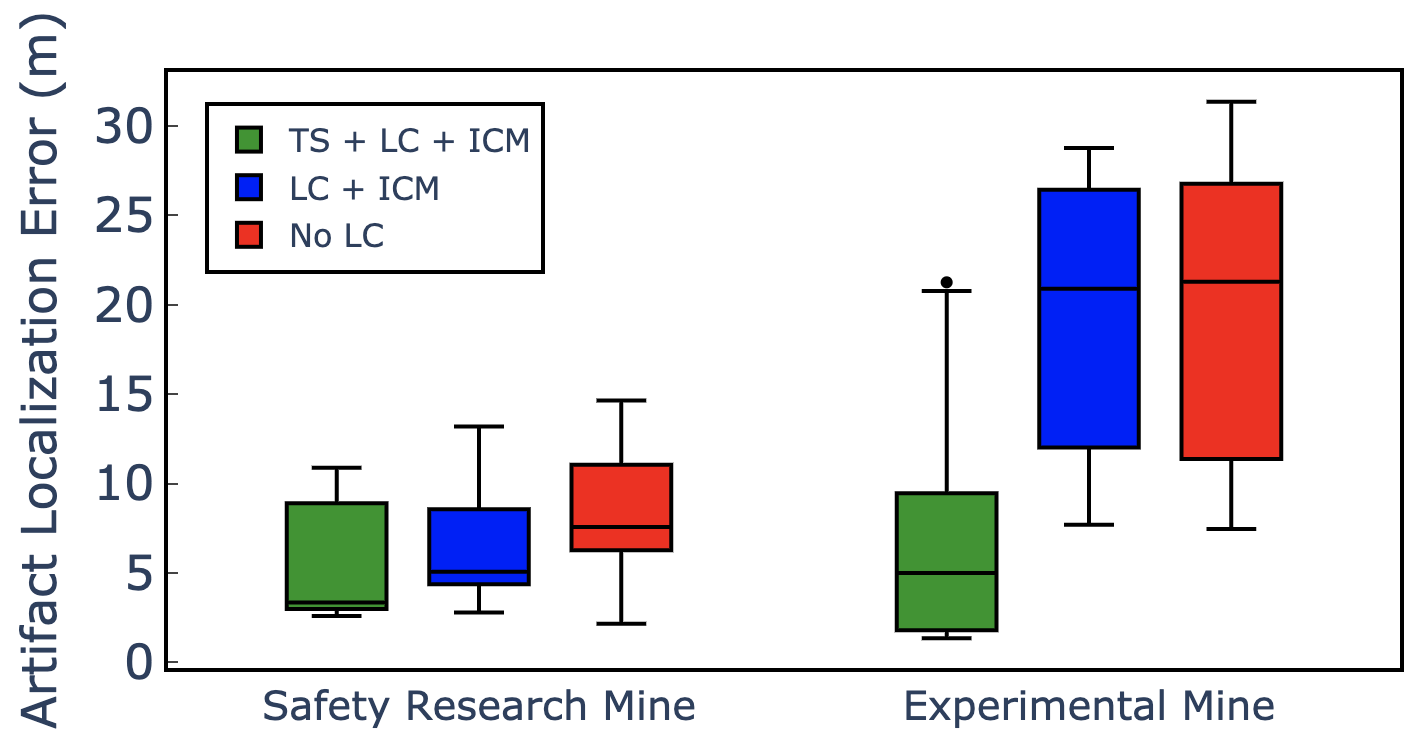}
	\caption{Artifact localization accuracy using LAMP, with and without loop closures (LC) and \iPCM, and with total station (TS) measurements in the Bruceton 
	Safety Research and Experimental 
	mines during the Tunnel Circuit of the DARPA Subterranean Challenge.  \label{fig:stats}\vspace{-2mm}}
\end{figure}

We show the performance and accuracy of LAMP in the Bruceton Safety Research and Experimental mines in Fig.~\ref{fig:stats}.
The boxplots show the artifact localization error with and without loop closures (LC) and outlier rejection (\iPCM). In addition, the localization performance achieved in the tunnel circuit of the DARPA challenge is shown in green; the latter shows a reduced error due to the inclusion of total station measurements of the robot pose in the pose graph.
\subsubsection{Multi-robot Mapping Evaluation}
\begin{figure}[t!]   
\centering 
  \includegraphics[width=1.0\columnwidth]{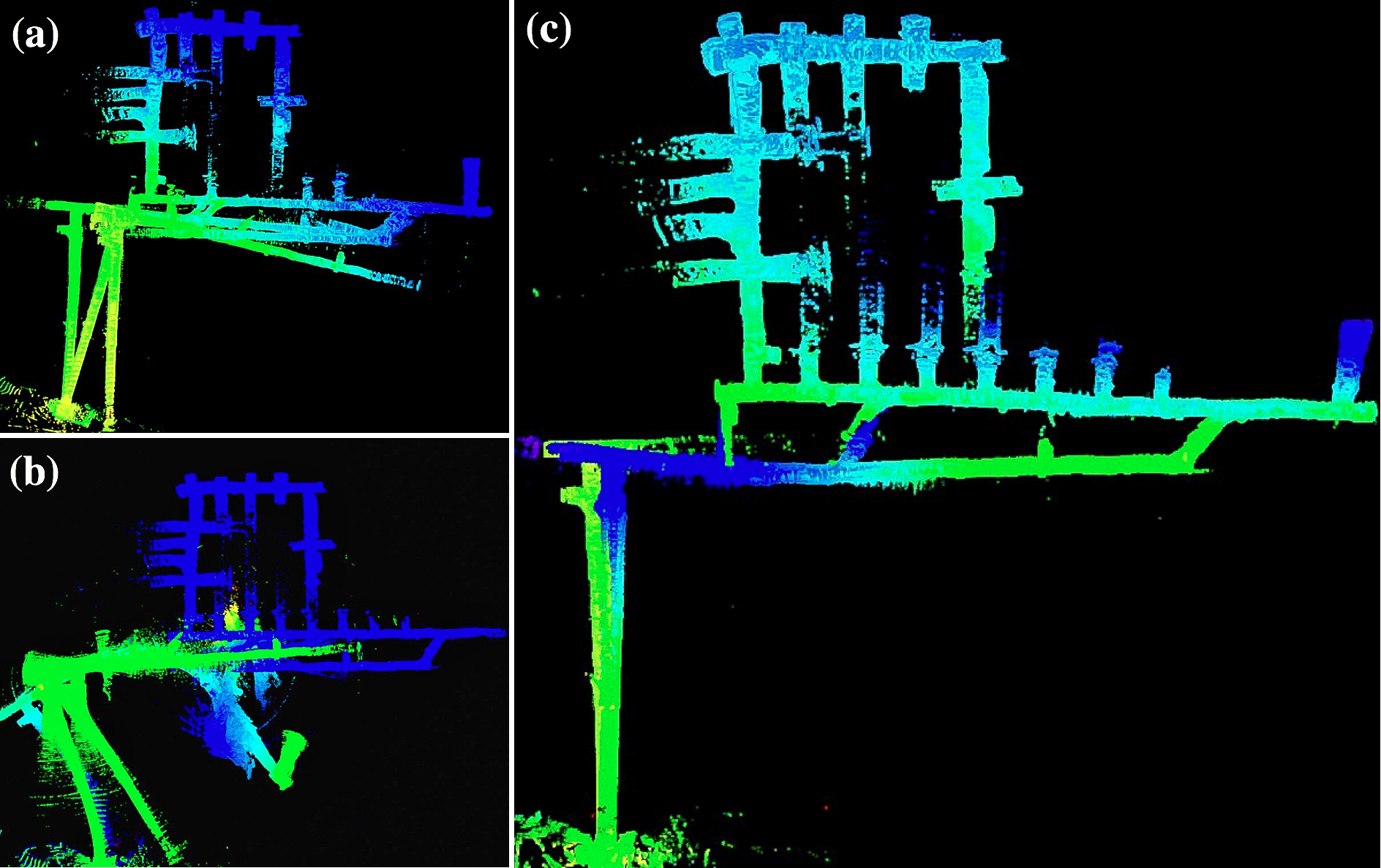}
  \caption{Multi-robot mapping experiment with two robots at the Bruceton Safety Research mine. (a) Base station map without performing inter-robot loop closure; 
  (b) base station map after inter-robot loop closure detection is enabled (no \iPCM);
  (c) base station map after inter-robot loop closure detection and \iPCM are enabled.\label{fig:multi}\vspace{-8mm}}
\end{figure}

To demonstrate our multi-robot architecture, we show the mapping results obtained at the base station. For this experiment, we set the \iPCM translation and rotation thresholds to 0.05 m and 0.005 rad. 
In order to reject outlying inter-robot loop closures the GICP is initialized with robot odometry and a loop closure is confirmed if the computed fitness score is greater than a set threshold (0.18). 
Fig.~\ref{fig:multi} shows the fused map from two robots (a) without loop closures, (b) with loop closures but no \iPCM, and (c) with \iPCM. We observe that without loop closures, not only the drift in the single robot trajectory estimates becomes apparent, but also errors in the initial calibration 
(based on the fiducial markers) lead to a significant divergence between the trajectories of the two robots.
If we enable loop closure detection, but we do not use \iPCM, a few incorrect loop closures visibly distort the map. With loop closures and \iPCM, the final map obtained is much improved. 
Fig.~\ref{fig:multi}(c), however, shows that \iPCM is not able to detect and reject an incorrect loop closure made at the first turn in the tunnel.

\section{Discussion}
\label{sec:discussion}
In this paper we presented LAMP, a centralized multi-robot SLAM system, developed in the context of the DARPA Subterranean Challenge, which provides a robust estimate of the trajectories of multiple robots and constructs a point cloud map using 3D lidar data. The system also includes a vision front-end to detect and localize objects of interest.

While LAMP was instrumental for the CoSTAR team~\cite{CoSTAR} to achieve second place in the tunnel circuit of the DARPA Subterranean Challenge, 
it is merely the result of the first phase of the competition, and, as such, it is a work in progress.
We believe there are several aspects that can be improved: some of these aspects can be addressed using more advanced techniques in the literature, while others expose fundamental open research problems. We review the main outstanding technical challenges  and future work below.

\ph{Lidar Odometry} Our choice of the lidar front-end odometry worked well in nominal conditions; however, 
as shown in Fig.~\ref{fig:frontend-drift-qual-eval}(b),(d) it might underestimate the robot motion in long symmetric corridors without detectable geometric features. In these scenarios, the scan matching problem become intrinsically ambiguous and GICP is easily trapped in local minima. This degeneracy is a well-known problem in lidar-SLAM and can be mitigated by fusing information from other sensors (e.g., wheel odometry, vision~\cite{ORB-SLAM, Khattak1}, IMU~\cite{Gentil, LION}), or adopting more recent methods that attempt to explicitly model the degeneracy~\cite{Kaess19}.

\ph{Loop Closure Detection}
While using the odometry initialization in GICP leads to missing many loop closure opportunities, the initialization proposed in this paper has the advantage of computing a large number of putative loop closures. Indeed, we are able to capitalize on loop closure opportunities despite a large amount of odometric drift. The downside is that our approach also produces a large amount of outliers. 
A natural evolution of the proposed system is to use feature-based approaches for robust global scan matching that do not require an initial guess~\cite{RobustFeatureRegistration}. Another practical solution might be to leverage other sensing modalities, e.g., use vision to detect loop closures~\cite{ORB-SLAM}. In addition, computation of precise covariances for loop closures and odometry is an avenue for future work.

\ph{Outlier Rejection} While we discussed \iPCM as a pragmatic solution to reject outlier loop closures (based on~\cite{PCM}),  
we believe that more research is needed to obtain truly robust SLAM solutions. 
\iPCM currently requires extensive parameter tuning and it often fails to strike a good balance between rejecting outliers and keeping inliers. This is expected, due to the approximate nature of the pairwise consistency checks. Recent approaches~\cite{Lajoie19ral} attempt to simultaneously assess the consistency of all loop closures accepted as inliers, but they currently do not scale well to large problems. The development of a robust global solver for PGO would increase the reliability of SLAM and relieve the user from time-consuming and problem-specific parameter tuning. 

\ph{Multi-robot Mapping}
 LAMP is currently a centralized SLAM system. While a centralized system works well for a small team of robots, we observed communication bottlenecks when performing tests with more than four robots. The literature offers fully distributed PGO solutions to reduce the communication burden~\cite{LucaCarlone, CommunicationAndSharing, Cunningham, Schuster, Vidal-Calleja}; however, these are not robust to outliers, which is a key requirement in our context.
 For underground exploration, it is also crucial to exchange a map representation and possibly provide an interpretable visualization to the user. 
 Point clouds are relatively cumbersome models and exchanging them (even after downsampling) quickly becomes impractical.
 Therefore, it is crucial to use map compression techniques (e.g.,~\cite{Zlot,OMeadhra19ral}) to reduce the communication bandwidth. In general, it would be desirable to have an automatic approach to adjust the type and amount of exchanged data depending on the available bandwidth. More compact and uncertainty-aware representations \cite{CRM} that can enable accurate risk prediction for motion planning are another direction for future research.

\clearpage
\renewcommand{\baselinestretch}{0.94}

\end{document}